\documentclass[10.5pt,letterpaper,twocolumn]{article} 
\usepackage[a4paper, total={6.3in, 10in}]{geometry}
\usepackage{graphicx}
\usepackage{authblk}
\usepackage{hyperref}
\usepackage{amsmath} 
\usepackage{amssymb} 
\usepackage[super,comma,sort&compress]{natbib}

\usepackage[utf8]{inputenc}   
\usepackage[T1]{fontenc}      
\usepackage{textcomp}         
\usepackage{lineno}
\usepackage{capt-of}
\usepackage[dvipsnames]{xcolor} 
\usepackage{nth}
\usepackage{soul}
\usepackage{xparse}
\UseRawInputEncoding

\usepackage{amsmath,amsfonts,bm}

\def\eqref#1{equation~\ref{#1}}

\def\1{\bm{1}}

\def\vb{{\bm{b}}}

\def\vw{{\bm{w}}}

\def\vy{{\bm{y}}}

\def\mW{{\bm{W}}}
\def\mX{{\bm{X}}}

\DeclareMathAlphabet{\mathsfit}{\encodingdefault}{\sfdefault}{m}{sl}
\SetMathAlphabet{\mathsfit}{bold}{\encodingdefault}{\sfdefault}{bx}{n}

\title{How much neuroscience does a neuroscientist need to know?}

\date{}

\author[1*]{James C.R. Whittington}
\author[2]{William Dorrell}
\affil[1]{Department of Experimental Psychology, University of Oxford}
\affil[2]{Gatsby Computational Neuroscience Unit, University College London}

\affil[*]{correspondence: \texttt{james.whittington@psy.ox.ac.uk}}

\begin{document}

\maketitle

\section*{Abstract}

How much of the brain's learned algorithms depend on the fact it is a brain? We argue: a lot, but surprisingly few details matter. We point to simple biological details---e.g. nonnegative firing and energetic/space budgets in connectionist architectures---which, when mixed with the requirements of solving a task, produce models that predict brain responses down to single-neuron tuning. We understand this as details constraining the set of plausible algorithms, and their implementations, such that only `brain-like' algorithms are learned. In particular, each biological detail breaks a symmetry in connectionist models (scale, rotation, permutation) leading to interpretable single-neuron responses that are meaningfully characteristic of particular algorithms. This view helps us not only understand the brain's choice of algorithm but also infer algorithm from measured neural responses. Further, this perspective aligns computational neuroscience with mechanistic interpretability in AI, suggesting a more unified approach to studying the mechanisms of intelligence, both natural and artificial.

\section*{Introduction}

\begin{figure*}[!t]
\begin{center}
\includegraphics[width=0.9\textwidth]{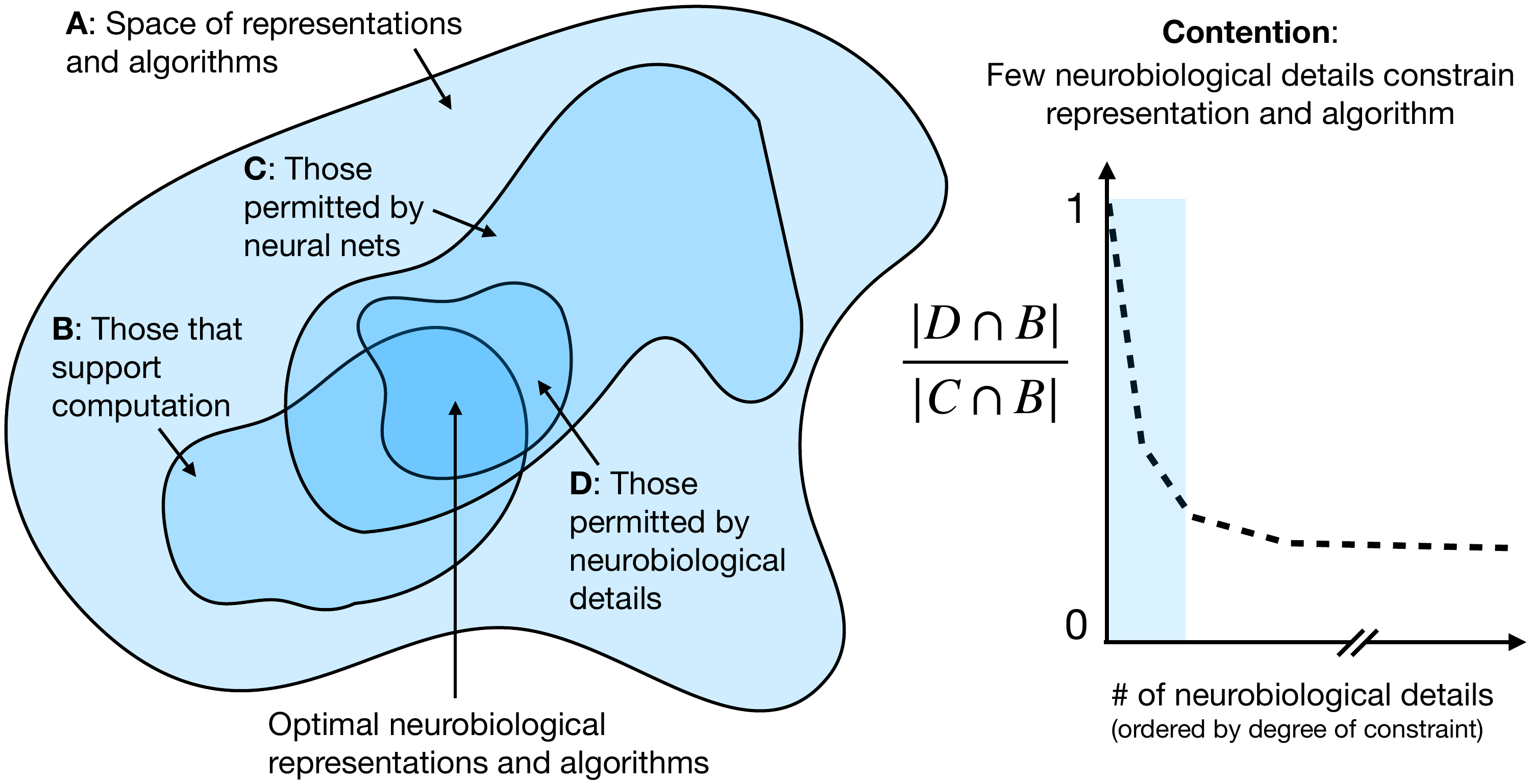}
\end{center}
\caption{\textbf{Hypothesis that few biological details constrain the space of effective algorithms.} 
\textbf{Left:} The space of potential algorithms taking place in the brain is constrained by computation and biological details.
\textbf{Right:} We contend that, conditioned on being a neural network, relatively few further biological details constrain the optimal algorithm.
}
\label{fig:AlgoSpace}
\end{figure*}

Neuroscience is often framed in levels\cite{marr_vision_1982, dayan_theoretical_2001}, with Marr's computational, algorithmic\footnotemark, and implementation the most famous. These levels are interdependent; task and behaviour (computational) determine algorithm, which require biological implementation. In this article we focus on the reverse: how lower level practicalities constrain the permissible set of algorithms (Figure \ref{fig:AlgoSpace}). Understanding which details constrain (and how) helps us to understand the brain's choice of algorithm and let us infer the brain's algorithm from the measured neural implementation.

\footnotetext{Sometimes known as `representation and algorithm', but we use `algorithmic' to avoid confusion with neural representations}

First, definitions. \textit{Algorithms} are sets of steps to solve problems by manipulating data within a particular format. Brains \textit{implement} these using neurons (and synapses) not machine code and CPUs. Understanding implementation, then, is understanding how and why neurons (and their connections) behave and how these processes are supported and maintained\footnotemark.

\footnotetext{We focus on the mechanics of brain networks rather than synaptic learning algorithm.}

But which details of biological implementation actually \textit{constrain} the algorithms a brain could use? While some details do constrain---e.g., local learning make exact backpropagation implausible\cite{crick_recent_1989, whittington_approximation_2017, whittington_theories_2019}---others seem optimal for \textit{serving} any connectionist algorithms: ion channel redundancy enables robustness of electrophysiological properties to temperature\cite{alonso_temperature_2020, alonso_visualization_2019}; interneurons types stabilise dynamics\cite{keijser_optimizing_2022}; synaptic learning rule diversity for stable learning\cite{confavreux_memory_2025}. As an analogy to computers, how the silicon is doped is vital, but it doesn't constrain the implementation of arbitrary algorithms; silicon \textit{serves}. Finite memory, on the other hand, \textit{constrains} what algorithm can be implemented.

Biological details abound---dynamics of ion channels and membrane potentials drive action potentials; development instils network priors; glia and astrocytes chaperone correct neural functioning; dendritic arbors integrate inputs---yet, there have been \textbf{surprising successes of single-neuron connectionism using few biological details}. Examples include the Marr-Albus-Ito circuitry for cerebellum and related structures\cite{marr_theory_1969, ito_neurophysiological_1970, albus_theory_1971}; reinforcement learning (RL) models of dopamine\cite{schultz_neural_1997}; continuous attractor network models (CANN) of heading direction\cite{skaggs_model_1995}; efficient coding models\cite{olshausen_emergence_1996}. These models feature no spike-time coding, few inhibitory neurons, no ion channel distributions or electrophysiology, no glia, no Dale's law, no dendrites, just point neurons and synaptic weights. The fact these models work suggests that you can get a surprisingly useful, and single-neuron, understanding using few details. Further, these models are structurally like task-optimised artificial neural networks (ANNs)---another exceptionally successful class of neural models (reviewed later)---suggesting understanding the brain's algorithm may parallel mechanistic interpretability ANNs.

Thus, this article contends that understanding the brain's algorithms and implementations down to a single neuron level may depend on relatively few biological details. This is not to dismiss biology's importance, but to highlight that one can go surprisingly far by considering 1) top-down constraints from tasks and behaviour; 2) `middle' constraints from different classes of neural architecture; 3) bottom-up biological details that break neural symmetries. Though unlikely to be universally true, this approach may both help parse biological complexity into logical components and provide a means of inferring implemented algorithm from neural tuning, providing a defence of building a single neuron understanding of the brain.

\section*{Top-down understanding of task and behaviour is critical but not enough}

Understanding the statistics of tasks and behaviour is fundamental to understanding intelligence\cite{marr_vision_1982, krakauer_neuroscience_2017} since algorithms are a reflection of the structures we see and use in the world. Understanding from Cognitive Science, Psychology, and Ethology have constrained the vast space of potential algorithms, and has led to a rich understanding of those used by the brain, including schemas, rule learning, complementary learning systems, path-integration, latent learning, bounded rationality, RL, uncertainty, or the Bayesian brain hypothesis.

However, turning these insights into mechanistic understanding is difficult due to limited constraints on hypothesis classes. Notably, many successful examples, e.g., ring attractors and striatal RL, concern simple well-defined algorithms (path-integration just integrates velocity). In contrast, many domains---like vision and language---lack such precise characterisations. 

Nevertheless modelling has made steady progress from symbolic\cite{turing_icomputing_1950, simon_human_1971}, connectionist\cite{rumelhart_parallel_1986}, and dynamical systems models\cite{thelen_dynamic_1994, kelso_dynamic_1995} to Bayesian\cite{tenenbaum_how_2011} and deep learning models\cite{yamins_performance-optimized_2014, kriegeskorte_deep_2015}. Now, our best cognitive models are often neural networks\cite{peterson_using_2021, binz_foundation_2025} and, while we rarely understand the implemented algorithm of ANNs, their learned algorithms can be very different to classical models, e.g., performing modular arithmetic in Fourier space\cite{nanda_progress_2023}, or language via analogical structures rather than idealised linguistic\cite{lindsey_biology_2025}; connectionism places a large constraint on the types of algorithms brains can learn.

\section*{Knowing about neural networks and behaviour might be nearly enough}

ANNs don't just sometimes learn similar algorithms to brains, but they appear to implement them with similar single neuron properties, e.g, in visual\cite{yamins_performance-optimized_2014, kriegeskorte_deep_2015, long_mid-level_2018}, auditory\cite{singer_sensory_2018}, prefrontal\cite{mante_context-dependent_2013, wang_prefrontal_2018, whittington_tale_2024} cortices, and the hippocampal formation\cite{cueva_emergence_2018, whittington_tolman-eichenbaum_2020, whittington_relating_2022}. Further, these models can be used predictively, e.g., to design stimuli that maximally excite specific neurons\cite{walker_inception_2019}, and can they recover the brain's precise mechanism, e.g., RNNs trained to path-integrate learn the same CANNs\cite{sorscher_unified_2023} found in the fly central complex\cite{kim_ring_2017}, and mammalian entorhinal cortex\cite{vollan_leftright-alternating_2025}.

This correspondence is remarkable and suggests many biological details (i.e. those not captured by ANNs) often serve the neural algorithm rather than constrain it. But two challenges remain. First, understanding what algorithm ANNs have learned is hard and requires careful analysis---controlled tasks, ablations, and causal manipulations\cite{zhao_explainability_2024}---mirroring neuroscience techniques. Further, ideally we'd have theory to relate task and chosen algorithm, beyond post-hoc interpretations. Second, many ANN architectures are universal function approximators and so any task could be solved via every algorithm imaginable; why should an ANNs choose the same solution as the brain? Indeed, in linear ANNs the network function and network configuration are disassociated\cite{braun_not_2025} (network similarity without functional similarity, and vice versa)---many network configurations solve the task.

Yet empirically, trained ANNs don't just learn any-old solution; instead they often match neural data, suggesting they use similar algorithms. This is likely due to implicit simplicity biases which, while poorly understood, favour biological solutions. Indeed, when the above linear ANN is trained with weight regularisation, the dissociation disappears and network configuration is unique to function.

Supporting this, recent theories have combined connectionist and basic biological constraints to recover phenomena related to the brain's algorithm. For example, task statistics bound neural manifold dimension\cite{gao_theory_2017, stringer_high-dimensional_2019}; place cells optimally tile manifolds under similarity matching objectives\cite{sengupta_manifold-tiling_2018}; 
shared neural pathways are optimal in gated linear networks\cite{saxe_neural_2022};
attractor manifolds are optimal in path-integrating RNNs\cite{dorrell_actionable_2023}; prefrontal slot-based attractor manifolds are optimal in RNNs solving structured sequence memory tasks\cite{whittington_tale_2024, dorrell_efficient_2025, piwek_recurrent_2023}. Importantly, network architecture places a critical constraint on which algorithms get learned, e.g., on arbitrary sequence memory tasks, RNNs learn slot attractors like those in prefrontal cortex, while RNNs with an external memory (e.g., a Hopfield Network) learn classic path-integrator like those in the hippocampal formation\cite{whittington_tale_2024}.

As such, it seems that connectionism with some constraints can tell us about the brain's algorithm. But what are these constraints? And how is the algorithm implemented at the single neuron level? Indeed, many of the above theories inform us about manifolds not single neurons. This is because there are symmetries at the level of neurons---e.g., rotation, scaling, permutation---that don't change the neural algorithm or underlying computation at the manifold level, but do change how individual neurons behave. Recent work however, both empirical and theoretical, is beginning to show that biological constraints also break these symmetries and reliably recapitulate single neurons coding\cite{dorrell_actionable_2023, sorscher_unified_2023, whittington_disentanglement_2023, margalit_unifying_2024}. We now highlight a few biological constraints that have proven powerful in our understanding of single neuron tuning.

\section*{Bottom-up biological details break symmetries of neural implementation}


\begin{figure*}[!t]
\begin{center}
\includegraphics[width=0.9\textwidth]{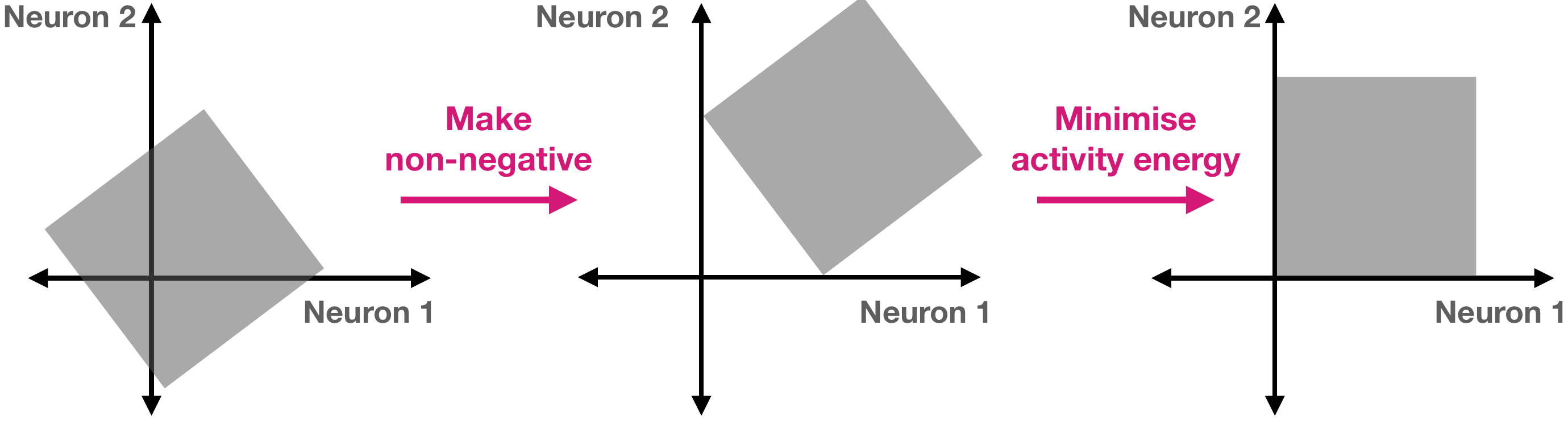}
\end{center}
\caption{\textbf{Intuition for how biological details can constrain neural response.} 
Two uniformly distributed independent factors represented with two entangled neurons (left). The neural population can be made nonnegative at the expense of activity energy (middle). Activity energy is minimised under a nonnegativity (and variance) constraint when the neurons are axis aligned to task factors (i.e. disentangled, right).
}
\label{fig:Disentanglement}
\end{figure*}

\textit{Breaking scale symmetry with energetic constraints}: 
The brain is energy-efficient, yet spikes and large synaptic connections are energetically expensive\cite{harris_synaptic_2012}. The efficient coding hypothesis\cite{rosenblith_possible_1961} posits that brains represent variables with minimal energy use, explaining tuning curves across brain regions\cite{olshausen_emergence_1996, lewicki_efficient_2002, simoncelli_natural_2001}. In machine learning, this energy minimisation is called regularisation. While the regularisation the brain uses is unknown, it is likely related to the number of spikes (firing rate)\footnotemark and connection strength. These constraints break scale symmetry and all but L2 regularisation break rotational symmetry.

\footnotetext{We note that baseline firing may prevent ion buildup, serving as as a protective mechanism at the cost of energy efficiency\cite{chintaluri_metabolically_2023}.}

\textit{Breaking rotational symmetry with single neuron constraints}: 
It is believed that firing rate convey information\cite{london_sensitivity_2010}, however rate cannot be negative. Nonnegativity breaks rotational symmetry and limits neural population activity. Indeed, ReLU activations (zero threshold) in ANNs produce more brain-like and modular neural responses\cite{nayebi_explaining_2021, yang_task_2019, driscoll_flexible_2024, whittington_disentanglement_2023}.

To intuit how these details shape tuning, consider two neurons coding for two variables (Figure \ref{fig:Disentanglement}). Under nonnegativity and energy efficiency, optimal coding assigns each neuron to just one variable---disentanglement, or modularity---commonly observed in the brain, e.g., `functional cell types' such as grid, band, and object vector cells. Importantly, these constraints interact with task statistics; modularity emerges only when input distributions have sufficient `square-ness'\cite{dorrell_range_2025}, explaining when grid cells warp towards rewards\cite{dorrell_range_2025} or when prefrontal slots are orthogonal to one another\cite{dorrell_efficient_2025}. Furthermore, when combined with task structure, these constraints explain single neuron tuning such as grid cell hexagonality (and their modularity)\cite{sorscher_unified_2023, dorrell_actionable_2023, whittington_disentanglement_2023, dordek_extracting_2016}.

\textit{Breaking permutation symmetry with neuron specific constraints}: 
Neurons differ in constraints, breaking permutation symmetry and encouraging clustering of similarly tuned units. A key example is wiring length minimisation\cite{rivera-alba_wiring_2011, chen_wiring_2006, zhang_universal_2000, chklovskii_wiring_2002}; long axons are space (and energy) expensive. Adding such constraints in ANNs yields realistic visual topologies\cite{doshi_cortical_2023, margalit_unifying_2024} and pinwheel maps\cite{koulakov_orientation_2001}

Another major source of permutation symmetry breaking is genetic cell types, e.g., dopamine neurons in RL\cite{schultz_neural_1997}; D1/D2-expressing medium spiny neurons in direct/indirect (go/no-go) pathways\cite{gerfen_d1_1990}; neurons only release glutamate (excitatory) or GABA (inhibitory) neurotransmitters (Dale's law). Though some functional modularity, that appears genetically constrained, may emerge regardless due to nonnegativity and energetic constraints. Furthermore, genetic precoding may be useful for effective embryology of structures involved in critical processes such as RL, action selection, sensory processing, cortical column. Regardless, these biological constraints support optimal algorithm (e.g. RL), while shaping single neuron tuning.

\begin{figure*}
    \centering
    \includegraphics[width=0.75\linewidth]{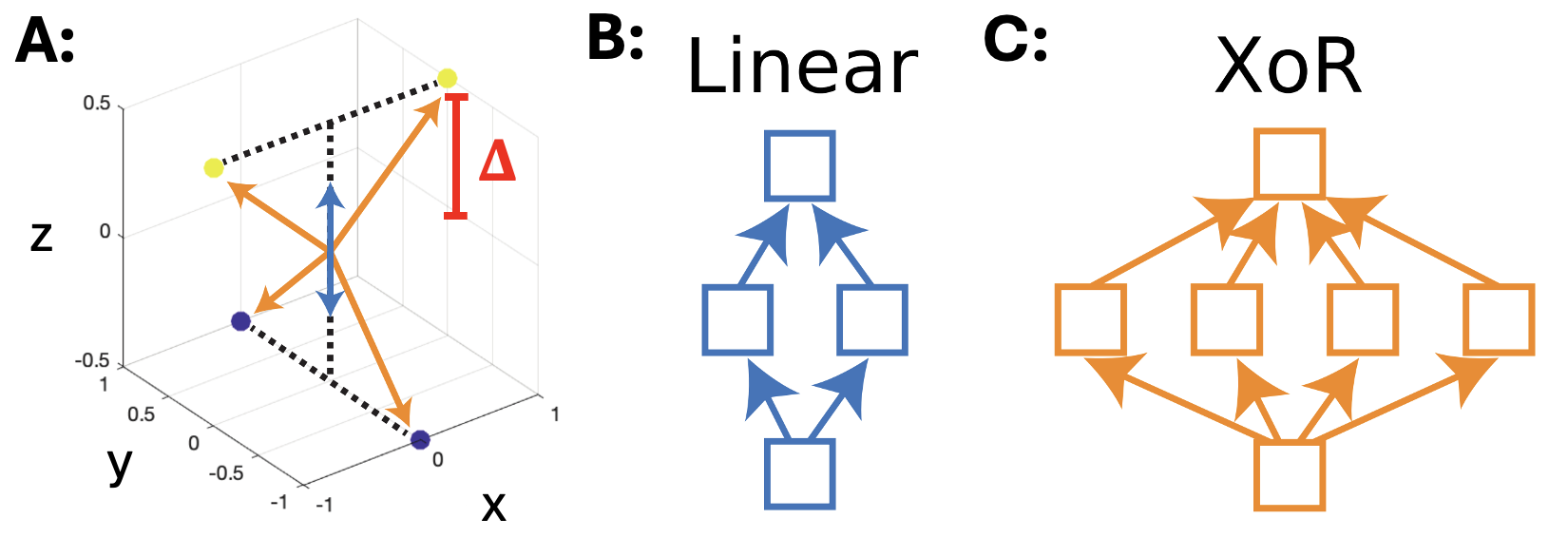}
    \caption{\textbf{Twisted XOR} \textbf{A:} The four 3D datapoints in the twisted XOR task. In the \textit{(x,y)} plane this is the classic XOR problem, whereas the \textit{z} direction simply encodes the label. $\Delta$ measures the size of the $z$ direction. There are two viable solutions, \textbf{B:} using two neurons to map the \textit{z} direction directly to labels, or \textbf{C:} using one neuron per datapoint to solve the task as an XOR. Which is learnt depends on the size of $\Delta$. Figure from the original work of Jarvis et al.\cite{jarvis_make_2025}.}
    \label{fig:Twisted_XOR}
\end{figure*}

\section*{A defence of the single neuron}

We've seen that biological constraints not only alter network-level manifolds but also systematically sculpt single-neuron responses. But if task, behaviour, and neural manifolds tell us much about the brain's algorithm, and algorithms are implemented by populations of neurons, then what's the point of understanding single neurons?

One answer is that when biological constraints 1) limit the space of algorithms and 2) shape single neuron tuning, then single neuron tuning can become aligned closely enough with the algorithm that it becomes diagnostic of the algorithm itself.

Consider the following synthetic example\cite{jarvis_make_2025}---an XOR task, but with a small twist; the input is not just an $(x,y)$ coordinate, but $(x,y,z)$ in which the $z$ dimension linearly separates the classes by a small distance $\Delta$ (Figure \ref{fig:Twisted_XOR}). One-hidden layer ReLU networks learn different algorithms based on the size of $\Delta$; when $\Delta$ is large the ANN uses the linear separability (the hidden layer represents the 4 data-points in 2 neurons). However, when $\Delta$ is small the ANN makes use of the nonlinearity (the hidden layer represents the 4 points in 4 separate neurons). A change in task structure leads to a change in algorithm. This was explained as a `race' between how fast each algorithm gets learned in each situation. However, in Appendix \ref{app}, we show the same result occurs due to energetic costs of the solutions differing with different $\Delta$. Here, biological constraints determine which algorithm gets learned (i.e., more than just the implementation of algorithm) which is then reflected in single neuron tuning; looking at the single neurons makes it very clear what algorithm is going on.

Indeed single neuron tunings, that has been shaped by biological constraints, have been critical in determining both the brain's algorithm and its particular implementation. Dopamine neurons helped us realise the brain was using temporal difference RL\cite{schultz_neural_1997}. D1 \& D2 neurons told us about action selection\cite{gerfen_d1_1990}. Simple and complex cells\cite{hubel_receptive_1962} helped us understand ConvNets\cite{fukushima_neocognitron_1980}, or centre-surround receptive fields taught us about efficient retinal coding\cite{atick_towards_1990, atick_what_1992}. The presence of modules of both spatial and conjunctive grid cells tells us about path-integration\cite{mcnaughton_path_2006}, leading to CANN models that fit the behaviour of both grid cells and fly ring attractors\cite{kim_ring_2017}. More speculatively, single neuron tuning across prefrontal cortex is often surprisingly tuned to abstract variables, like hierarchical concepts of structure\cite{shima_categorization_2007}, conceptual actions like reversing a sequence\cite{ohbayashi_conversion_2003}, or progress\cite{el-gaby_cellular_2024}. Further, in the last example single neuron tuning was vital to decoding the proposed neural algorithm.

The modular tuning of these neurons is the product of biological constraints. Without these constraints we would only have a manifold level understanding, and it's not clear whether any of these inferences about algorithm would have gotten any easier (likely much harder).

Indeed confusion can arise from considering manifolds alone. Grid and place cells are clearly different functional cell types, but their respective low dimensional projections (e.g., PCA on activity as a rodent explores a 2D arena) appear very similar. Knowing the single neurons not only led to the development of different CANN models (implementation) for the different cell types, but it led to the understanding that the cells serve different purposes: place cells for memory\cite{manns_evolution_2006, eichenbaum_hippocampus_1999} and grid cells for learning generalisable spatial primitives\cite{behrens_what_2018,whittington_tolman-eichenbaum_2020}. This is slightly a straw man: for place and grid cells there are other analyses---e.g., remapping---that would cause difference in the PCA plots. But our argument is not that single neurons are the only way to succeed. Instead, their behaviour provides useful constraints for reasoning about the implemented algorithm, so why not use them?

\section*{Discussion}

We have contended that our eventual understanding of the brain---at the algorithmic level---will rely on an understanding of the interaction between tasks and behaviours, neural networks, and biological details that limit the space of algorithms and break symmetries at the implementation level. With algorithm heavily constrained by task, behaviour, neural network architecture, and biological details, and implementation further constrained by biological details. This view aligns the neuroscientist with the AI mechanistic interpretability researcher who aims to build a circuit level understanding of ANNs and perhaps suggests that cheap interrogation of ANNs could replace some costly animal experimentation. Interestingly, the biological details we have found particularly important---nonnegativity and energy efficiency---facilitate interpreting ANNs\cite{bricken2023monosemanticity}, suggesting the brain is a more interpretable neural network than often thought\cite{barack_two_2021, chung_neural_2021, langdon_unifying_2023}. 

What biological details likely don't break symmetries of algorithm? This is hard to predict and we don't want to put our necks on the line, but if pushed we'd posit ion channels and their dynamics, spiking neural networks, plasticity rules, dendrites, glia, and intracellular proteins and signalling pathways, gene expression and plasticity machinery, EI balance, oscillations play a lesser role in most cases we know so far. Further we suggest other details like local microcircuits, cortical layers, neuromodulation, neurotransmitters do not shape algorithm, but do shape implementation. There are always exceptions: neurons sometimes compute with spike timings not rates\cite{stopfer_impaired_1997, stopfer_short-term_1999}; neurons can have meaningfully different electrophysiological events\cite{yang_purkinje-cell_2014, muller_complex_2023}; dopamine tags memories in synapses to govern memory replay\cite{mcnamara_dopaminergic_2014, atherton_memory_2015}. However we still contend that, with just a few biological details---e.g., nonnegativity and energy efficiency---one can go surprisingly far in understanding much of the brain's chosen algorithm. As such, including these details in neural network models is a must for computational neuroscientists. 

Our thesis is a long way from substantiated. Convincingly showing that just a few choice biological details interact with computation to constrain neural algorithm requires not only actually knowing what the computation is, but also being able to decipher the neural algorithm. Much more work is required in cases like vision where we don't know the underlying algorithm or often even the task, as opposed to more cognitive areas such as hippocampus and prefrontal cortex where much of our argument has been focused. 

Nevertheless, while our eventual understanding of algorithms may depend on few biological details, theorists should know as many details as possible. Experiments run on details (e.g., optogenetics, protein tagging) and we need experiments for theory verification and exploration as otherwise we won't ever get to that `eventual' understanding. So theorists, buckle up and learn some neuroscience.

\section*{Acknowledgements}

We thank Kris Jensen, Chen Sun, Tim Muller for helpful feedback on the manuscript, and Devon Jarvis for conversations about twisted XOR. We thank the following funding sources: Sir Henry Wellcome Postdoctoral Fellowship (222817/Z/21/Z) to JCRW.; European Research Council Starting Grant (NARFB/101222868; JCRW); the Gatsby Charitable Foundation to WD.

\clearpage

\bibliography{references}

\clearpage

\begin{appendix}
\onecolumn

\section{Twisted XOR Energy Calculation}\label{app}
Jarvis et al. 2025 consider the following task, a variant of the XOR task. There are 4 $(x, y)$ datapoints of inputs and labels:
\begin{equation}
    \mX = \begin{bmatrix}
        1 & 1 & -1 & -1 \\
        1 & -1 & 1 & -1 \\
        \Delta & -\Delta & -\Delta & \Delta
    \end{bmatrix}
    \qquad \vy = \begin{bmatrix}1 & -1 & -1 & 1\end{bmatrix}
\end{equation}
We see that if $\Delta=0$ then this is the classic XOR task. When considering a one-hidden layer ReLU network with no bias trained on this task, \citet{jarvis_make_2025} pose two solutions, a `linear' one that attends only the $z$ direction of the input (termed linear as it uses the linear separability of the datapoints based on the $z$ direction and so only requires $2$ effective neurons in the hidden layer), and a `non-linear' one in which the hidden layer neurons are tuned to single inputs (termed `non-linear' as it does not utilise the linear separability of the datapoints based on the $z$ direction and so requires $4$ effective neurons in the hidden layer). They show that the speed at which the two solutions are learned differ, and vary as $\Delta$ varies. For very small $\Delta$, according to the neural race reduction\cite{saxe_neural_2022}, the linear solution learns slower than the non-linear solution, and so the resultant network is non-linear. Conversely if $\Delta$ is high enough the linear solution learns quicker, and so the final network is linear. The transition between the two occurs at $\Delta = \sqrt{\frac{2}{3}}$.

Rather than considering learning speeds, we instead ask the question, if the network weights are regularised, is one solution energetically favoured over another and how does this change with $\Delta$?

\subsection{Calculation}

To analyse this, we calculate the L2 weight losses of the two networks as a function of $\Delta$. We'll call weights in the first layer $\mW \in \mathbb{R}^{N \times 3}$ (with columns $\vw_i$) and in the second layer $\vb \in \mathbb{R}^{N}$ where $N$ is the number of effective neuron in the hidden layer.

\subsubsection{The Linear Solution}

The linear solution is effectively just two neurons in the hidden layer\footnotemark, with the following set of weights:
\begin{equation}
    \vw_1 = \begin{bmatrix}
        0 \\ 0 \\ \alpha
    \end{bmatrix} \quad \vw_1 = \begin{bmatrix}
        0 \\ 0 \\ -\alpha
    \end{bmatrix} \quad \vb = \begin{bmatrix}
        \beta \\ -\beta
    \end{bmatrix}
\end{equation}

\footnotetext{We need two neurons because, without bias, one neuron has to encode the positive and the other the negative part of the inputs. Even with a bias, however, the two neuron solution is preferred both energetically and by the `neural race'.}

Thus after applying the ReLU activation, the activity in the hidden layer is either $\begin{bmatrix} \alpha\Delta , 0 \end{bmatrix}$ or $\begin{bmatrix} 0 , \alpha\Delta \end{bmatrix}$ depending on which datapoint was inputted. And so, in order to fit the data the following equation must be satisfied:
\begin{equation}
    \beta\alpha\Delta = 1
\end{equation}
And subject to this constraint we seek to minimise the weight loss:
\begin{equation}
    \mathcal{L}_W = ||\mW||^2_F + ||\vb||^2_F = 2(\alpha^2 + \beta^2)
\end{equation}
The optimal solution to this constrained optimisation problem (i.e., using Lagrange multipliers) is:
\begin{equation}
    \alpha = \beta = \frac{1}{\sqrt{\Delta}} \qquad \mathcal{L}_W = \frac{4}{\Delta}
\end{equation}

\subsubsection{The Non-Linear Solution}
The alternative is to have four classes of neurons in your hidden layer, each pointing towards one of the datapoints. Let's consider just one of them for now ($vw_1$), and get the others by symmetry.
\begin{equation}
    \vw_1 = \frac{\alpha}{\sqrt{2+\Delta^2}}\begin{bmatrix}
        1 \\ 1 \\ \Delta
    \end{bmatrix}  \quad \vb = \begin{bmatrix}
        \beta \\ -\beta \\ -\beta \\ \beta
    \end{bmatrix}
\end{equation}
Thus, assuming $\Delta$ is small, after applying the ReLU activation, the activity in the hidden layer is a permutation of $\begin{bmatrix} \sqrt{2+\Delta^2}\alpha , 0 , 0, 0\end{bmatrix}$ depending on which datapoint was inputted. And so, in order to fit the data the following equation must be satisfied:

\begin{equation}
    \sqrt{2+\Delta^2}\alpha\beta = 1
\end{equation}
And subject to this we need to minimise:
\begin{equation}
    \mathcal{L}_W = ||\mW||^2_F + ||\vb||^2_F  = 4(\alpha^2 + \beta^2)
\end{equation}
The solution to this constrained optimisation problem is:
\begin{equation}
    \alpha = \beta= (2 + \Delta^2)^{-\frac{1}{4}}\qquad \mathcal{L}_W = 8(2 + \Delta^2)^{-\frac{1}{2}}
\end{equation}

\subsubsection{Comparison}

Setting the two losses equal to each other we can derive that the transition point is:
\begin{equation}
    \Delta = \sqrt{\frac{2}{3}}
\end{equation}
This is the same critical point as originally derived by Jarvis et al., but using a different argument.
\end{appendix}

\end{document}